\documentclass[twocolumn,showpacs,prl,amsmath,amssymb,floatfix]{revtex4}
\usepackage{dcolumn}
\usepackage{amsmath,amssymb}
\usepackage{bm}
\usepackage{epsfig}
\newcommand{\bd}{\bm}
\begin{document}
\title{Spectral function and quasi-particle damping of interacting bosons in two dimensions} 
\author{Andreas Sinner$^1,$ Nils Hasselmann$^2$, and Peter Kopietz$^1$}
\vspace{2mm}  
\address{$^1$Institut f\"{u}r Theoretische Physik, Universit\"{a}t
    Frankfurt, Max-von-Laue-Strasse 1, 60438 Frankfurt, Germany\\
	$^2$International Center for Condensed Matter Physics, Universidade de Bras\'{\i}lia, 
Caixa Postal 04667, 70910-900 Bras\'{\i}lia, DF, Brazil}
%
\date{November 4, 2008}
\begin{abstract}
\noindent
We employ the functional renormalization group to
study dynamical properties of the two-dimensional Bose gas. Our approach is free of infrared divergences,
which plague the usual diagrammatic approaches, and is consistent
with the exact Nepomnyashchy identity, which states that the 
anomalous self-energy vanishes at zero frequency and momentum. 
We recover the correct infrared behavior of the propagators and present explicit results for the spectral line-shape, 
from which we extract the quasi-particle dispersion and damping.
\end{abstract}

\maketitle

The excitation spectrum of the weakly interacting Bose gas 
has been studied with field theoretical methods for more than half a century 
\cite{Bogoliubov47}
and is qualitatively well understood: at high energies the excitations resemble
those of free bosons whereas at low energies the excitations are collective 
Goldstone modes which result from the symmetry broken ground state \cite{Hugenholtz59,Shi98,Andersen04}.
In $D=3$, this picture has recently been confirmed experimentally using the Bragg spectroscopy 
technique on cold atoms~\cite{Ozeri05}. However,
while perturbative approaches (based on a renormalized $T$-matrix) to the 
weakly interacting Bose gas have been successfully applied to study
the excitation spectra and the damping of quasi-particles \cite{Beliaev58}, the self-energies
in these approaches violate the exact
Nepomnyashchy identity \cite{Nepomnyashchy75}, which states that the anomalous
self-energy at zero frequency and momentum vanishes. Moreover, the perturbative
approach is plagued by infrared (IR) divergences \cite{Shi98,Andersen04}, which makes 
a numerical evaluation non-trivial already at second order in the effective coupling constant.
 Modern renormalization group (RG)
methods \cite{Pistolesi04,Dupuis07,Wetterich07} have resolved this problem and recovered the correct low-energy structure of
the self-energies which are in accordance with the Goldstone character of the
excitations and the  Nepomnyashchy identity.
The calculations in Refs.~\cite{Pistolesi04,Dupuis07,Wetterich07} were however limited to the
asymptotic regime near zero energy and are thus unable to describe both the damping of quasi-particles and the 
crossover from the low energy Goldstone-like dispersion to the free particle dispersion at higher
energies. 

There is a renewed interest in the physics of interacting bosons since 
the transition from weak to strong interaction can now be studied experimentally 
via the Feshbach resonance technique \cite{Papp08}.
The excitations of the weakly interacting
gas are qualitatively different from those of strongly interacting liquids such
as  $^4$He and the connection between the two remains elusive.  
A consistent strong coupling approach, which would allow a description
of the strongly interacting gas starting from a microscopic model is at 
present not available.
In two dimensions, the need to go beyond
mean field theory is more urgent, since the $s$-wave scattering
length vanishes and the usual expansion parameter, the gas parameter, is replaced
by a new emergent parameter~\cite{Schick71}. 
In this Letter we show how the
functional RG (FRG) can be employed to calculate 
the single-particle spectral density
of the interacting Bose gas in $D$ dimensions and present numerical results for $D=2$.
Our approach can be extended to include arbitrarily strong
{\em  local} correlations and might offer a possible route to the physics of
strongly interacting bosons.

We consider bosons with mass $m$, chemical potential $\mu$ and a repulsive contact interaction $u_0$ at zero temperature. The bare Euclidean action is
\begin{eqnarray}
S[\bar\psi,\psi]  =  \int d^Dx d \tau\left[\bar{\psi}^{}(\partial_\tau-\frac{{\bd{\nabla}}^2}{2m}-\mu) \psi^{}
+ \frac{u_0}{2} (\bar{\psi} {\psi})^2\right],\,
\label{eq:InAct}
\end{eqnarray}
where the spatial integrals should be regularized by means of a short-distance cutoff $\Lambda^{-1}_0$, which is related to the finite extent of the interaction or, for the hard core bosons, to the size of particles. Note that the model~ (\ref{eq:InAct}) depends on two dimensionless parameters $\tilde \mu_0=2m\mu\Lambda^{-2}_0$  and $\tilde u_0=2mu_0\Lambda^{D-2}_0$, which give the chemical potential and interaction energy in units of $\Lambda^2_0/2m$. 
The FRG is conveniently formulated in terms of the generating functional
of irreducible vertex functions $\Gamma_\Lambda[\bar\phi,\phi]$, depending
on an IR cutoff $\Lambda$ which eventually will be removed~\cite{Founders}.
The functional $\Gamma_\Lambda$ is defined through $\Gamma_\Lambda[\bar\phi,{\phi}]={\cal L}_\Lambda[\bar\phi,\phi]+\int_K \bar{\phi}_K 
G^{-1}_{0,\Lambda}(K)\phi_K$ where  $K=({\bm k},i\omega)$ and
$\int_K = (2\pi)^{-(D+1)}\int d^D k ~ d \omega$ is the integration over momenta and
Matsubara frequencies. Here,
${\cal L}_\Lambda[\bar\phi,\phi]$ is the Legendre transform of the generating functional
of connected Green functions and $G_{0,\Lambda}^{-1}=i\omega - \epsilon_{k}+\mu - R_\Lambda(k)$ is the inverse free propagator with dispersion $\epsilon_{k}=k^2/2m$.
The regulator function $R_\Lambda(k)$ will be specified below. 
We shall use the following {\em non-local} potential approximation for $\Gamma_\Lambda$,
\begin{equation}
\Gamma_\Lambda
=\int_K \bar{\phi}_K \sigma_\Lambda(K)\phi_K 
+ \frac{1}{2} \int_K \delta\rho_K u_\Lambda(K) \delta\rho_{-K} \, .
\label{eq:defgamma}
\end{equation} 
The quantity $\delta\rho_K=\int_Q \bar{\phi}_{Q}\phi_{K+Q}
- \delta_{K,0} \rho_\Lambda^0$ is the Fourier-transform of $\rho_X-\rho_\Lambda^0$
where $X=({\bm x},\tau)$. Here, 
$\rho_X=|\phi_X|^2$ 
is the local density and  $\rho^0_{\Lambda}$ is
the flowing condensate density, which has a finite limit for $\Lambda\to 0$ in the symmetry broken state.
The non-local character of the potential coupling function $u_\Lambda(K)$,
which was not included in previous FRG approaches \cite{Dupuis07,Wetterich07},
is essential to describe correctly the low energy sector of the
theory.
Although we will in the end calculate the complete momentum and frequency dependence
of  $\sigma_\Lambda(K)$ and $u_\Lambda(K)$, we shall also employ
 the following low energy expansion,
\begin{subequations}
  \begin{eqnarray}\vspace{-1cm}
    \sigma_\Lambda(K)&\approx&  \mu (1-X_\Lambda) + i \omega (1-Y_\Lambda) 
\nonumber \\
 &+& \epsilon_{k}(Z_\Lambda^{-1}-1)
      - (i \omega)^2 V_\Lambda  \, , 
    \label{eq:approxsigma} \\
    \vspace{-1cm}    
    u_\Lambda(K) &\approx& u_\Lambda(0) \equiv u_\Lambda \, ,
    \label{eq:approxu}
  \end{eqnarray}
\end{subequations}
with $X_\Lambda=\rho^0_\Lambda u_\Lambda/\mu$.
The flow of the parameters $Y_\Lambda, Z_\Lambda,$ $V_\Lambda, \rho_\Lambda^0$, and $u_\Lambda$
will be determined self-consistently. The initial values can be read off from the action (\ref{eq:InAct}) and are given by $X_{\Lambda_0}=Y_{\Lambda_0}=Z_{\Lambda_0}=1$, $V_{\Lambda_0}=0$ and $u_{\Lambda_0}=u_0$. The~ initial~ condition~ for~ the~ condensate~ density is given by the mean field result $\rho_{\Lambda_0}^0=\mu/u_0$. 
The usual Beliaev Green functions are defined via the matrix equation 
${\bm G}_\Lambda^{-1}(K)={\bm G}_{0,\Lambda}^{-1}(K)-{\bm \Sigma_\Lambda}(K)$, where
\begin{eqnarray}
  {\bm G}_\Lambda(K) &=&\left(
    \begin{array}{cc}
      G_\Lambda^N(K) & G_\Lambda^A(K) \\
      G_\Lambda^A(K)^* & G_\Lambda^N(-K)
    \end{array}
  \right) \, ,   \\
  {\bm G}_{0,\Lambda}^{}(K)&=&\left(
    \begin{array}{cc}
      G_{0,\Lambda}^{}(K) & 0\\
      0 & G_{0,\Lambda}^{}(-K)
    \end{array}
  \right)\, ,  \\
  {\bm \Sigma}_\Lambda(K)&=&\left(
    \begin{array}{cc}
      \Sigma_\Lambda^N(K) & \Sigma_\Lambda^A(K) \\
      \Sigma_\Lambda^A(K)^* & \Sigma_\Lambda^N(-K)
    \end{array}
  \right)\, .
\end{eqnarray}
The normal $\left(\Sigma^N\right)$ and anomalous $\left(\Sigma^A\right)$ self-energies are obtained from an expansion of Eq.~(\ref{eq:defgamma}) in $\delta\phi_X=\phi_X-\phi_\Lambda^0$ \cite{Schuetz05,Sinner08a},
where we assume a real valued order parameter such that 
$\rho_\Lambda^0=\bar{\phi}^0_{\Lambda}\phi_\Lambda^0=
(\phi_{\Lambda}^0)^2$.  The self-energies 
are obtained from $\sigma_\Lambda(K)$ and $u_\Lambda(K)$ via
\begin{subequations}
  \begin{eqnarray}
    \Sigma_\Lambda^N(K) & = & \sigma_\Lambda(K)+ \rho_\Lambda^0 [u_\Lambda(0)+u_\Lambda(K)] \, , 
    \label{eq:approxsigman}\\
    \Sigma_\Lambda^A(K) & = & \rho_\Lambda^0 u_\Lambda(K) \, .
    \label{eq:approxsigmaa}
  \end{eqnarray}
\end{subequations}
Conversely, we may determine $\sigma_\Lambda(K)$, 
which at $T=0$ has a well defined low-frequency
and momentum expansion \cite{Nepomnyashchy75}, and the non-local coupling function
$u_\Lambda(K)$, which will become non-analytic for $\Lambda\to 0$, from a calculation of the
self-energies 
for {\em all} frequencies
and momenta.
The effective action (\ref{eq:defgamma}) elegantly divides the well behaved
parts of the self-energies from the non-analytic ones. It further guarantees, 
since only the flows of the $U(1)$-invariants 
$u_\Lambda$, $\rho_\Lambda^0$ and $\sigma_\Lambda$ are retained, that the Hugenholtz-Pines 
relation $\Sigma^N_\Lambda(0)-\Sigma^A_\Lambda(0)=\mu$ \cite{Hugenholtz59} is obeyed for
all $\Lambda$. 
We use the Litim  cutoff function  
$R^{}_\Lambda(k)=\left(1-\delta_{{\bm k},0}\right)(2mZ^{}_\Lambda)^{-1}
\left(\Lambda^2-k^2 \right)\Theta\left(\Lambda^2-k^2\right)$, which has good convergence properties \cite{Litim01}
and yields simple momentum integrals,
to derive the flow equations for the coupling parameters \cite{Dupuis07,Wetterich07}.
We define $\kappa_\Lambda=K_D[1-\eta^{z}_\Lambda/(D+2)]/D$ 
with $K_D=2^{1-D}\pi^{-D/2}/\Gamma[D/2]$, where $\eta^{z}_\Lambda=\Lambda \partial_\Lambda \ln Z_\Lambda$ is the anomalous dimension of the fields,  and $\Delta_\Lambda=u_\Lambda \rho_\Lambda^0$. We 
arrive at 
\cite{Dupuis07}
\begin{subequations}
  \begin{eqnarray}
    \partial_\Lambda \rho_\Lambda^0 &=& 
    4  \frac{\Lambda^{D+1}\kappa_\Lambda}{2m Z_\Lambda}\int \frac{d \omega}{2\pi}
    \sum_{n=0}^3\frac{c^{(\rho)}_{n} \omega^{2n}}{{\cal D}_\Lambda^2(i \omega)} 
    \label{eq:rhoFlow} \, , \\
    \partial_\Lambda u_\Lambda &=& 
    4 u_\Lambda^2  \frac{ \Lambda^{D+1}\kappa_\Lambda}{2mZ_\Lambda} \int \frac{d \omega}{2 \pi}\sum_{n=0}^3 
    \frac{c^{(u)}_{n} \omega^{2n}}{{\cal D}_\Lambda^3(i\omega)} \, , \label{eq:gFlow}
\end{eqnarray}
\end{subequations}
where
$  {\cal D}_\Lambda(i\omega)=Y_\Lambda^2 \omega^2+[\tilde{\epsilon}_\Lambda
  +V_\Lambda \omega^2] 
[\tilde{\epsilon}_\Lambda+V_\Lambda \omega^2 +2 \Delta_\Lambda] \, ,$
and $\tilde{\epsilon}_\Lambda=\epsilon_\Lambda/Z_\Lambda$.  Here we have defined 
$c^{(\rho)}_0 =  
\tilde{\epsilon}^2_\Lambda+\tilde{\epsilon}_\Lambda\Delta_\Lambda+\Delta^2_\Lambda$,
$c^{(\rho)}_1 =  V_\Lambda(2\tilde{\epsilon}_\Lambda+\Delta_\Lambda)-Y^2_\Lambda$,
and $c^{(\rho)}_2 = V^2_\Lambda$.
The flow of  $\rho_{\Lambda}^0$ is chosen such
that in the expansion of $\Gamma_\Lambda$ in the 
field $\delta \phi_X=\phi_X-\phi_\Lambda^0$ no terms linear in $\delta\phi_X$ appear \cite{Schuetz05}, i.~e., one always expands around the flowing
minimum of $\Gamma_\Lambda$. For the coefficients entering Eq.~(\ref{eq:gFlow}) we have
$c^{(u)}_0 =  5\tilde{\epsilon}^3_\Lambda+3\tilde{\epsilon}^2_\Lambda\Delta_\Lambda
+6\tilde{\epsilon}_\Lambda
\Delta^2_\Lambda+4\Delta^3_\Lambda$,
$c^{(u)}_1 =  3 V_\Lambda \left(5\tilde{\epsilon}^2_\Lambda
+2\tilde{\epsilon}_\Lambda\Delta_\Lambda+
2\Delta^2_\Lambda\right)-Y^2_\Lambda\left(7\Delta_\Lambda+11\tilde{\epsilon}_\Lambda\right)$,
$c^{(u)}_2 = V_\Lambda \left[3V_\Lambda\left(5\tilde{\epsilon}_\Lambda+\Delta_\Lambda\right)
  -11Y^2_\Lambda\right]$, and
$c^{(u)}_3 = 5 V^3_\Lambda$.
The flow equations of the derivative 
terms $Y_\Lambda$, $Z_\Lambda$ and
$V_\Lambda$~ are~ extracted~ from~ the~ flow~ equation~ of 
$\Sigma_\Lambda^N(K)-\Sigma_\Lambda^A(K)=\rho^0_\Lambda u_\Lambda+\sigma_\Lambda(K)$.
We find
\begin{subequations}
  \begin{eqnarray}
    \hspace{-.8cm}
    \partial^{}_\Lambda Y^{}_\Lambda &=& 
    -8 \rho^0_\Lambda u^{2}_\Lambda Y_\Lambda\frac{\Lambda^{D+1}\kappa_\Lambda}{2mZ^{}_\Lambda}
    \int\frac{d\omega}{2\pi} \sum_{n=0}^{2}\frac{c^{(y)}_{n}\omega^{2n}}
    {{\cal D}_\Lambda^3(i\omega)}  ,
    \label{eq:SFlow} 
    \\ \hspace{-.8cm}
    \partial^{}_\Lambda Z^{}_\Lambda &=& 4 \rho^0_\Lambda u^{2}_\Lambda  
    \frac{\Lambda^{D+1}K_D}{2mD}\int\frac{d\omega}{2\pi}\frac{1}{{\cal D}_\Lambda^2(i\omega)} \, ,
    \label{eq:ZFlow} 
    \\ \hspace{-.8cm}
    \partial^{}_\Lambda V^{}_\Lambda &=& 8 \rho^0_\Lambda u^{2}_\Lambda 
    \frac{\Lambda^{D+1}\kappa_\Lambda}{2mZ^{}_\Lambda}\int\frac{d\omega}{2\pi}\sum_{n=0}^2 
    \frac{c^{(v)}_{n}\omega^{2n}}{{\cal D}_\Lambda^3(i\omega)} \, ,
    \label{eq:VFlow}
  \end{eqnarray}
\end{subequations}
where
$c^{(y)}_0 =  \tilde{\epsilon}^2_\Lambda-2\tilde{\epsilon}_\Lambda\Delta^{}_\Lambda
-2\Delta^2_\Lambda$,
$c^{(y)}_1 =  Y^2_\Lambda+2(\tilde{\epsilon}_\Lambda-\Delta^{}_\Lambda)V^{}_\Lambda$,
$c^{(y)}_2 =  V^2_\Lambda$, 
and 
$c^{(v)}_0 = -Y^2_\Lambda(\tilde{\epsilon}_\Lambda+\Delta^{}_\Lambda)-\tilde{\epsilon}_\Lambda
(\tilde{\epsilon}_\Lambda+2\Delta^{}_\Lambda)V^{}_\Lambda$,
$c^{(v)}_1 = 2V^{}_\Lambda [Y^2_\Lambda+V^{}_\Lambda(\tilde{\epsilon}_\Lambda+\Delta^{}_\Lambda)]$,
$c^{(v)}_2 = 3V^3_\Lambda$.

\begin{figure}[h]
\centering
\includegraphics[height=3.2cm, width=5.5cm]{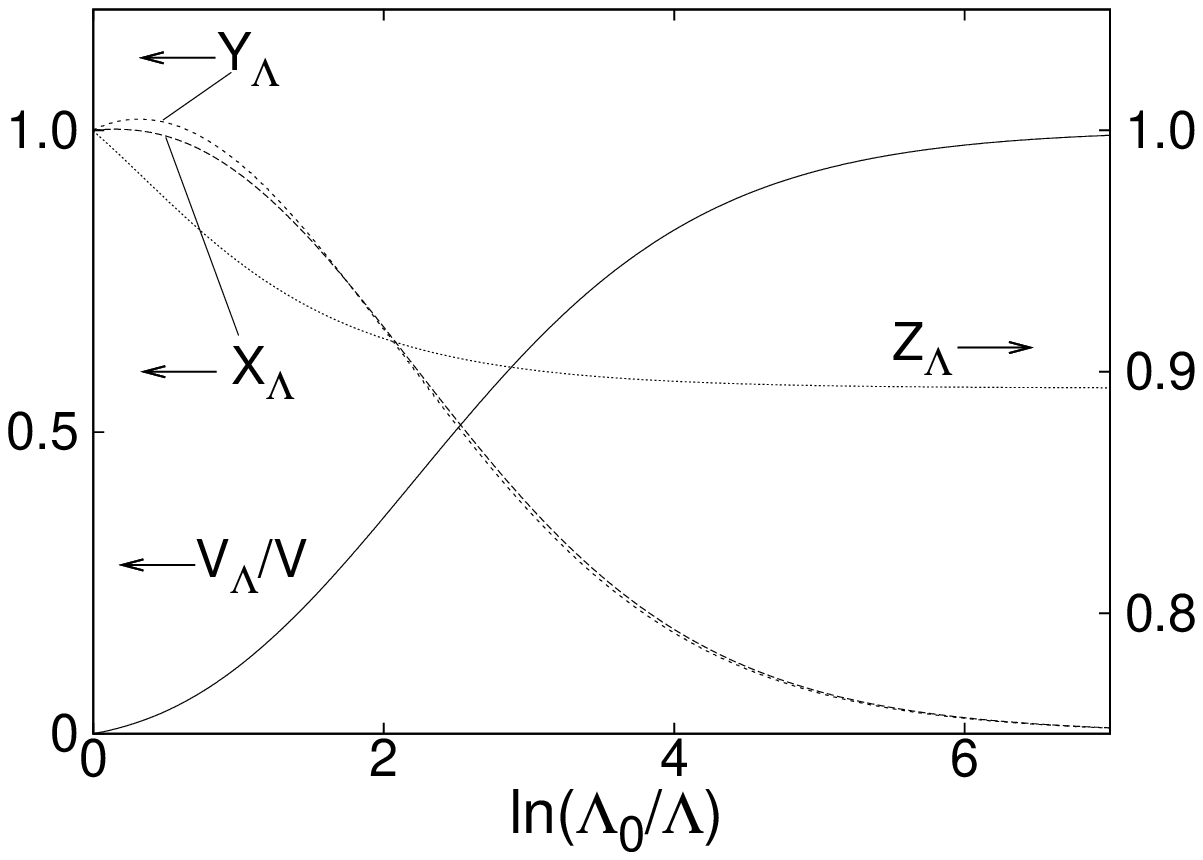}
\includegraphics[height=3.2cm, width=5.5cm]{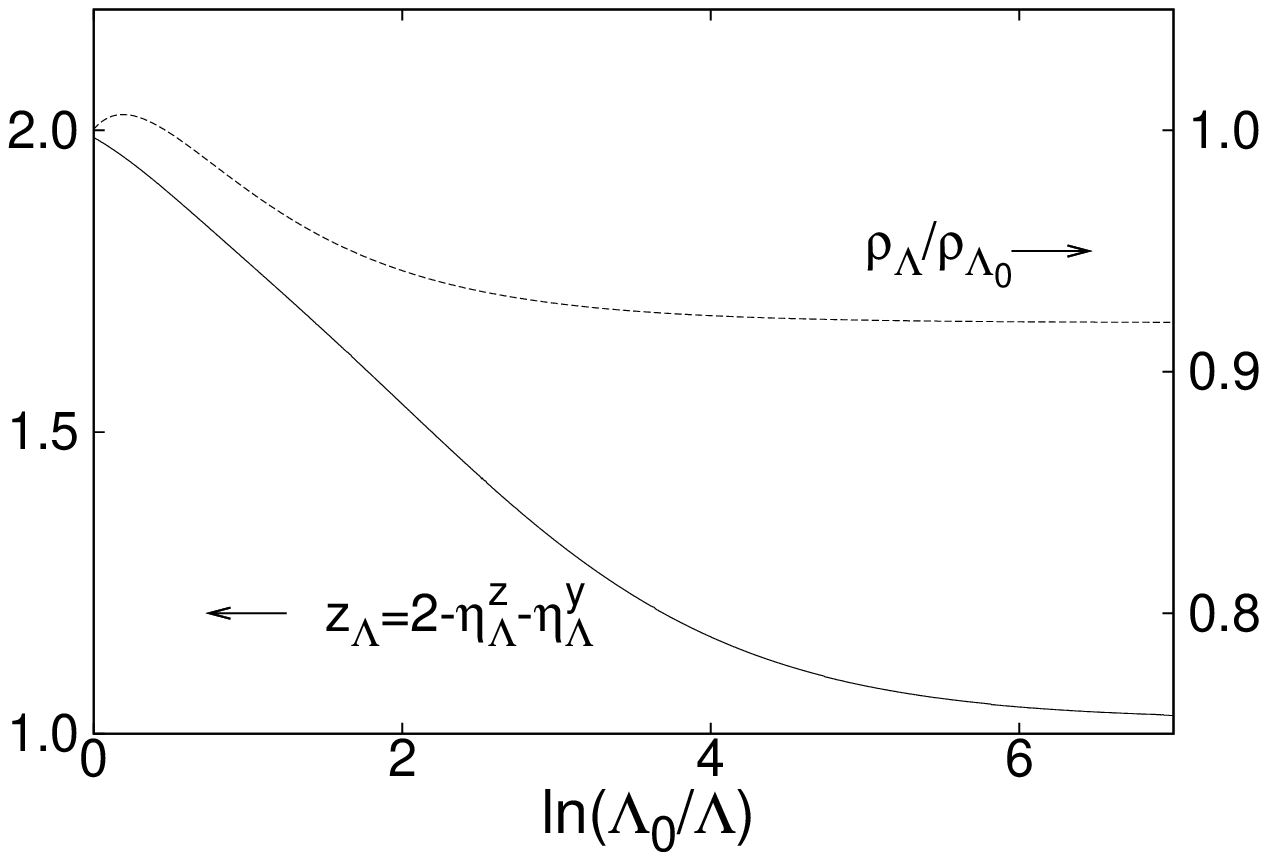}
\caption{Typical flows of the parameters $X_\Lambda$, $Y_\Lambda$, $Z_\Lambda$, $V_\Lambda$, $\rho^0_\Lambda$ and dynamical exponent $z_\Lambda = 2-\eta_\Lambda^z-\eta_\Lambda^y$. The curves are calculated for $\tilde \mu_0=2m\mu\Lambda^{-2}_0=0.4$ and $\tilde u_0=2mu_0=4$. The arrows point to the relevant scales.} 
\label{fig:dyn}
\end{figure}

Similar equations were already discussed in Ref.~\cite{Dupuis07,Wetterich07} 
and we briefly summarize the results in two dimensions.  At $T=0$ the condensate density $\rho_\Lambda^0$ flows
to some finite limit $\rho_{\Lambda=0}^0>0$, while
the coupling constant $u_\Lambda$ vanishes for $\Lambda \to 0$ which ensures
$\Sigma_{\Lambda\to 0}^N(K=0)=\mu$ and $\Sigma_{\Lambda\to 0}^A(K=0)=0$, in accordance
with both the Hugenholtz-Pines relation \cite{Hugenholtz59} 
and the Nepomnyashchy identity \cite{Nepomnyashchy75}. Furthermore,
$Y_\Lambda$ vanishes for $\Lambda\to 0$, again in accordance with
exact results \cite{Nepomnyashchy75}. Since $V_\Lambda$ acquires a finite 
value and $Z_\Lambda$ remains finite for $\Lambda\to 0$, the low energy modes are phonons
with a linear dispersion and velocity
$c=(2m V Z)^{-1/2}$ with $V=V_{\Lambda=0}$ and $Z=Z_{\Lambda=0}$. 
In Fig.~\ref{fig:dyn} we show flows of the coupling parameters and of  
the dynamical exponent $z_\Lambda = 2-\eta^{y}_\Lambda-\eta^{z}_\Lambda$ 
where $\eta_\Lambda^y=\Lambda\partial_\Lambda \ln Y_\Lambda$ 
and $\eta_\Lambda^z=\Lambda\partial_\Lambda \ln Z_\Lambda$, which conveniently illustrates 
the crossover from the free particle regime with $z=2$ to the Goldstone regime, where $z=1$. 
To work out how the crossover appears in the momentum dependence of
the self-energy, the truncations (\ref{eq:approxsigma},\ref{eq:approxu})
are however not sufficient. Moreover, within these truncations one obtains 
$G^A(K)=\lim_{\Lambda\to 0}G^A_\Lambda(K)\equiv 0$ which violates
the asymptotic $K\to 0$ relation ~\cite{Gavoret64,Nepomnyashchy75}
\begin{equation}
  G^N(K)\sim -G^A(K)\sim \frac{m \rho^0 c^2}{\rho}\frac{1}{(\omega^2+c^2 k^2)} \, ,
  \label{eq:limitGreen}
\end{equation}
where $\rho^0$ is the condensate density, $\rho$ the boson density and $c$ the thermodynamic
sound velocity. This relation can be satisfied only within an approach which takes account of
the non-analytic structure of $u_\Lambda(K)$  \cite{Nepomnyashchy75}.
We further note that the crossover energy scale $\Delta_c=4 \pi\rho^0 /[m \ln (\rho^0 a^2)]$, 
obtained by Schick   \cite{Schick71} for hard core bosons of diameter $a$, emerges from the $z=2$
regime of Eqs.~(\ref{eq:gFlow},\ref{eq:rhoFlow}) in the dilute limit and for large $u_{\Lambda_0}$.

We now proceed with the evaluation of the self-energies.
Truncating the momentum dependence of the irreducible
vertices on the right-hand-side of the flow equations 
in accordance with the truncation of $\Gamma_\Lambda$ in 
Eqs.~(\ref{eq:defgamma}) and  (\ref{eq:approxsigma},\ref{eq:approxu}), we find 
the flow equation of the self-energies
(the single scale propagators $\dot{G}_\Lambda^{A,N}(K)$ are defined
via $\dot{\bm G}_\Lambda(K)=-{\bm G}_\Lambda(K) \partial_\Lambda{\bm G}_{0,\Lambda}^{-1}(K) 
{\bm G}_\Lambda(K)$ ),
\begin{widetext}
  \begin{subequations}
    \begin{eqnarray}
      \partial_\Lambda\Sigma^{N}_\Lambda(K)& = & 2u^{}_\Lambda\int_{Q}\left\{ 
        \dot{G}^{N}_\Lambda(Q)+\dot{G}_\Lambda^{A}({Q})\right\} 
      \nonumber
      -  4u_\Lambda^2 \rho^0_\Lambda\int_{Q}\Big\{ \dot{G}_\Lambda^{N}({Q}) 
      \big[G_\Lambda^{N}(Q+K)+G_\Lambda^{N}(Q-K)+G_\Lambda^{N}(-Q+K)       
      \\
      & & + 2 G^{A}_\Lambda(Q-K) \big]
      +2\dot G_\Lambda^{A}(Q) 
      \big[G_\Lambda^{A}(Q+K) + G_\Lambda^{N}(Q+K) \big] \Big\} \, ,       
      \label{eq:NormVertFlow} 
      \\
      \partial_\Lambda\Sigma^{A}_\Lambda(K)& = & 2u_\Lambda\int_{Q}
      \dot{G}^{N}_\Lambda(Q) 
      -4 u_\Lambda^2 \rho^0_\Lambda \int_Q \Big\{ \dot{G}^{N}_\Lambda(Q) \big[G_\Lambda^{N}(Q+K)
      +G_\Lambda^{N}(Q-K)+G_\Lambda^{A}(Q+K)
      +G_\Lambda^{A}(Q-K) \big] \nonumber 
      \\
      && +  \dot{G}^{A}_\Lambda(Q) \big[G_\Lambda^{N}(Q+K)+G_\Lambda^{N}(Q-K)
      +3G_\Lambda^{A}(Q+K) \big] \Big\} \, .
      \label{eq:AnomVertFlow}
    \end{eqnarray}
  \end{subequations}
\end{widetext}

To solve Eqs.~(\ref{eq:NormVertFlow},\ref{eq:AnomVertFlow}),
we adopt a technique we applied previously to the classical $\phi^4$-model \cite{Sinner08a}
(see \cite{Ledowski04,Blaizot06} for approaches to classical models in their symmetric phase).
Instead of attempting to calculate a completely self-consistent solution, we use a
non-self-consistent approach where on the right-hand-side we approximate the self-energies
entering the Green functions using Eqs.~(\ref{eq:approxsigma},\ref{eq:approxu}) and
(\ref{eq:approxsigman},\ref{eq:approxsigmaa}). The full solution is expected to appreciably
deviate from the non-self-consistent solution only
for large momenta (at the order of $\Lambda_0$) or  at strong coupling. 
With Eqs.~(\ref{eq:approxsigma},\ref{eq:approxu}) and the already determined
flows of $u_\Lambda,\rho^0_\Lambda,Y_\Lambda,Z_\Lambda$ and $V_\Lambda$, we can simply
integrate Eqs.~(\ref{eq:NormVertFlow},\ref{eq:AnomVertFlow}) to obtain the complete
momentum and frequency dependence of the self-energies $\Sigma^{N/A}(K)=\lim_{\Lambda\to 0}
\Sigma_\Lambda^{N/A}(K)$. The resulting expressions for $\Sigma^{N/A}(K)$ are free 
from IR divergences since the coupling constant $u_\Lambda$ vanishes in the
IR limit. 

An important result of this approach is the correct description of the non-analytic structure
of $u_\Lambda(K)$ \cite{Nepomnyashchy75} 
with leading terms linear in  $|\omega|$ and $|{\bm k}|$  (for $D=2$). The Green functions
$G^N(K)$ and $G^A(K)$ are obtained from ${\bm G}^{-1}(K)={\bm G}_0^{-1}(K)-{\bm \Sigma}(K)$~\cite{Shi98}
and one can check that since $\rho^0 |u^{}_{\Lambda\to0}(K)|\gg|\omega^2 V +\epsilon_k/Z|$ for
small $K$, $G^N(K)$ and $G^A(K)$ have the limiting behavior given in Eq.~(\ref{eq:limitGreen}) 
if we identify \cite{Nepomnyashchy75}
$\rho^0/\rho =Z$ and $c=(2m V Z)^{-1/2}$.

\begin{figure}[h]
\centering
\includegraphics[height=3.9cm]{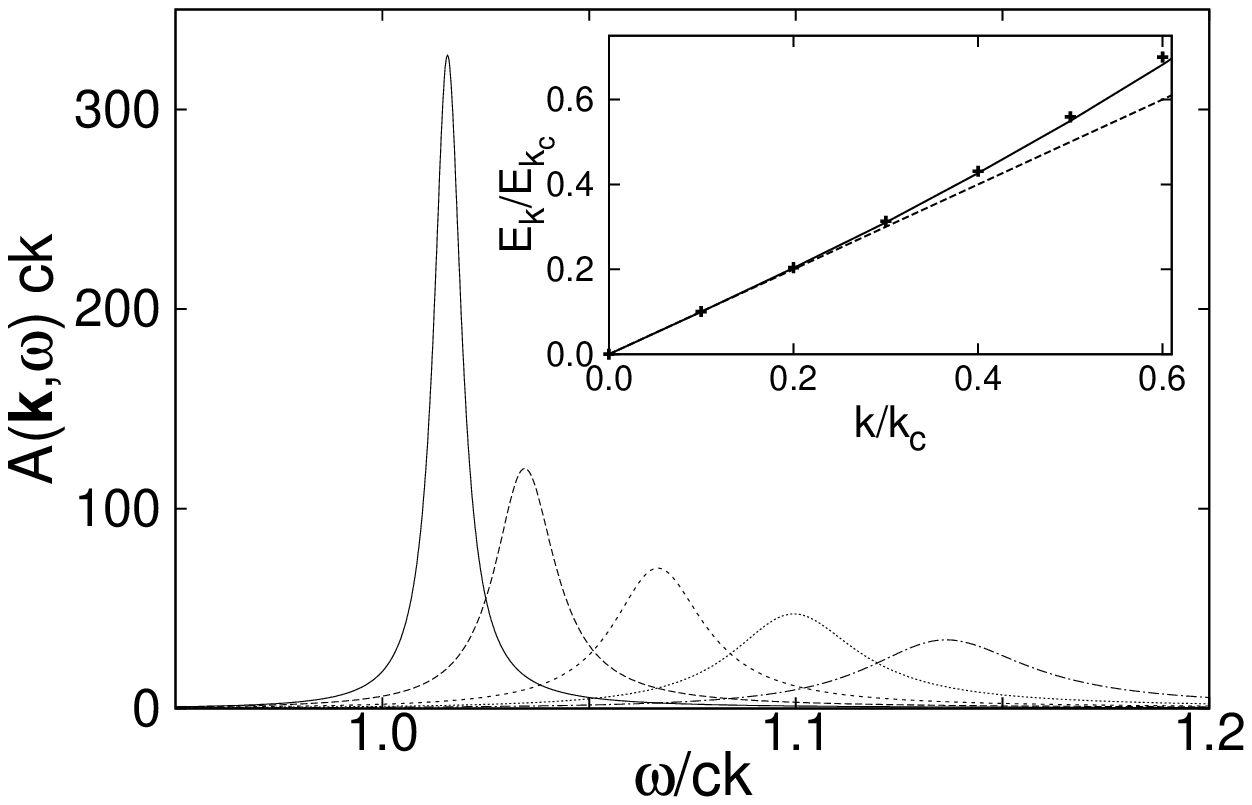}
\caption{Single-particle spectral density $A({\bm k},\omega)$ as a function of 
$\omega$, for different values of $k$, calculated for $\tilde \mu_0=0.15$ and $\tilde u_0=15$. The inset shows the quasi-particle dispersion $E_k$ which deviates at large $k$ from linearity but is well
described by the Bogoliubov form $E_k^2=\epsilon_k^2+ c^2 k^2$ with 
renormalized velocity $c$  (black dots). The peaks (from left to right) correspond to $k/k_c=0.2,$ $0.3, 0.4, 0.5$ and $0.6$, where $k_c=2mc$.}
\label{fig:specdens}
\centering
\includegraphics[height=3.9cm]{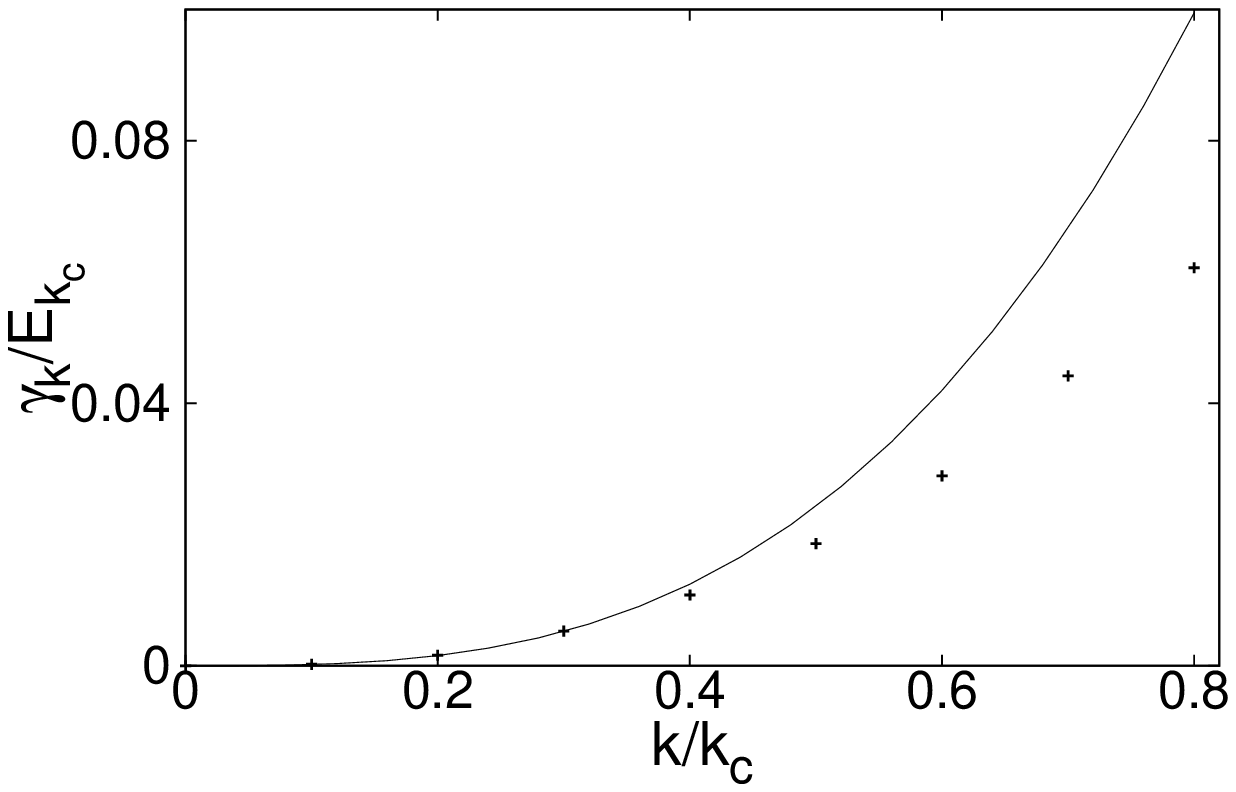}
\caption{FRG result for the quasi-particle damping $\gamma_k$. Black dots are extracted from the spectral density in  Fig.~\ref{fig:specdens}, while the solid line fits them as $\gamma_k\approx 0.194k^3/2mk_{c}$.}
\label{fig:qp-damp}
\end{figure}
The single-particle spectral density $A({\bm k},\omega)$ is obtained from the imaginary part of the normal propagator 
via $A({\bm k},\omega)=-2 \mbox{Im} G^N({\bm k}, i \omega \to \omega+i 0)$ and requires 
analytic continuation to real frequencies. We used the standard Pad\'e technique \cite{Vidberg77} with 200 points. 
In Fig.~\ref{fig:specdens}
the spectral density is plotted as a function of $\omega>0$ for different
momenta. One clearly observes a finite peak broadening, 
which grows with increasing momenta. The broadening arises from Beliaev damping \cite{Beliaev58}. 
Because of the upward curvature of the dispersion, a momentum- and energy-conserving
decay of  quasi-particles into pairs of quasi-particles is allowed. 
A second order perturbative analysis \cite{Kreisel08} shows that in $D$ dimensions this damping at small momenta
and for weak coupling has the form
\begin{equation}
\gamma^{(2)}_k\approx I^{}_{D}(2m\rho^0)^{-1}k^{3-D}_0 k^{2D-1},
\label{eq:damping}
\end{equation}
where $I^{}_D=2^{-4}3^{\frac{D+1}{2}} K_{D-1} \int_0^1dx(x-x^2)^{D-1}$,
$k_0=2mc_0$, and $c_0=\sqrt{\rho^0 u_0/m}$  is the mean field velocity.
While in $D = 3$ the perturbative expression is independent of $c_0$ and
reproduces the Beliaev result $\gamma^{(2)}_{k} = 3k^5/(640 \pi m \rho^{0})$, 
in $D=2$
the damping depends explicitly on $c_0$ and is thus expected to be renormalized. 
An analysis of the data  in  Fig.~\ref{fig:specdens} 
reproduces the $k^3$-behavior for small momenta as predicted by Eq.~(\ref{eq:damping}) and shown in Fig.~\ref{fig:qp-damp}. 
However, the prefactor $\alpha_0=I^{}_D(2m\rho^0)^{-1}k^{3-D}_0$ in Eq.(\ref{eq:damping}) should be replaced by a function $\alpha(\tilde \mu_0,\tilde u_0)$ of the relevant dimensionless parameters $\tilde \mu_0$ and $\tilde u_0$ of the model (\ref{eq:InAct}). For example for $\tilde \mu_0=0.4$ and $\tilde u_0=4$ we obtain $c/c_0\approx 1.01$ and $\alpha/\alpha_0\approx0.915$, while the results for $\tilde \mu_0=0.15$ and $\tilde u_0=15$ are  $c/c_0\approx0.669$ and $\alpha/\alpha_0\approx 0.526$.
The inset of Fig.~\ref{fig:specdens} shows the extracted quasi-particle 
dispersion, which is well fitted by
$E_k^2=\epsilon_k^2+c^2 k^2$ with renormalized velocity $c$.
The deviation of the spectrum from linearity occurs at the same momentum where the damping deviates from the cubic asymptote. 

In conclusion, we have presented FRG results for the spectral function of interacting bosons in $D=2$ 
in an approach which is consistent with the Nepomnyashchy identity $\Sigma^A(0)=0$ and the Hugenholtz-Pines relation
$\Sigma^N(0)-\Sigma^A(0)=\mu$. 
While previous RG calculations~\cite{Pistolesi04,Dupuis07,Wetterich07} were limited to the Goldstone regime, our approach allows calculating the entire spectral line-shape, including the dispersion and the damping of the quasi-particles. Our truncation captures the non-analytic structure of the self-energies, described by the non-local potential coupling $u_\Lambda(K)$, which is essential for a correct description of the
low-energy physics. We are currently investigating extensions of our method to include arbitrarily strong {\em local} correlations. This would possibly provide a non-perturbative access to strongly interacting bosons. 

We thank A. L. Chernyshev for detailed discussions
on Beliaev damping, H. O. Jeschke for providing us with the Pad\'e routine 
and C. Eichler for a correction. 
Support by the SFB/TRR49 and by a DAAD/CAPES PROBRAL grant is acknowledged.


\vspace{-0.5cm}
\begin{thebibliography}{99}
%
\bibitem{Bogoliubov47} N. N. Bogoliubov, {J. Phys. USSR} {\bf 11}, 23 (1947).
%
\bibitem{Shi98} H. Shi and A. Griffin, {Phys. Rep.} {\bf 304}, 1 (1998).
%
\bibitem{Andersen04} J.~O.~Andersen, Rev.~Mod.~Phys. {\bf 76}, 599 (2004).
%
\bibitem{Hugenholtz59} N. M. Hugenholtz and  D. Pines, {Phys. Rev.} {\bf 116}, 489 (1959).
%
\bibitem{Ozeri05} R.~Ozeri, N.~Katz, J.~Steinhauer, and N.~Davidson, Rev.~Mod.~Phys.~{\bf 77}, 187 (2005).
%
\bibitem{Beliaev58} S. T. Beliaev, {JETP} {\bf 7}, 289 (1958); {\em ibid.} {\bf 7}, 299 (1958).
%
\bibitem{Nepomnyashchy75} A. A. Nepomnyashchy and Yu. A. Nepomnyashchy, {JETP Lett.} {\bf 21}, 
1 (1975) and {JETP} {\bf 48}, 493 (1978); 
Yu. A. Nepomnyashchy, {JETP} {\bf 58}, 722 (1983).
%
\bibitem{Pistolesi04} 
C. Castellani, C. Di Castro, F. Pistolesi, and G. C. Strinati, {Phys. Rev. Lett.} {\bf 78}, 
1612 (1997); 
F. Pistolesi, C. Castellani, C. Di Castro, and G. C. Strinati, {Phys. Rev.} B {\bf 69}, 024513 (2004).
%
\bibitem{Dupuis07} N.~Dupuis and K.~Sengupta, {Europhys. Lett.} {\bf 80}, 50007 (2007).
%
\bibitem{Wetterich07} C.~Wetterich, {Phys. Rev.} B {\bf 77}, 064504 (2008); 
S.~Floerchinger and C.~Wetterich, Phys.~Rev.~A {\bf 77}, 053603 (2008); Phys.~Rev.~A {\bf 79}, 013601 (2009) 
%
\bibitem{Papp08} S.~B.~Papp, J.~M.~Pino, R.~J.~Wild, S.~Ronen, C.~E.~Wieman, D.~S.~Jin, and E.~A.~Cornell,
Phys.~Rev.~Lett.~{\bf 101}, 135301 (2008).
%
\bibitem{Schick71}  M. Schick, {Phys. Rev.} A {\bf 3}, 1067 (1971). 
%
%
\bibitem{Founders} C. Wetterich, {Phys. Lett.} B {\bf 301}, 90 (1993); T. R. Morris, {Int. J. Mod. Phys.} A {\bf 9}, 2411 (1994).
%
\bibitem{Schuetz05}  F. Sch\"{u}tz and P. Kopietz, {J. Phys.} A {\bf 39}, 8205 (2006).
%
\bibitem{Sinner08a} A. Sinner, N. Hasselmann, and P. Kopietz, 
{J. Phys.: Cond. Mat.} {\bf 20}, 075208 (2008).
%
\bibitem{Litim01} D. F. Litim, {Phys. Rev.} D {\bf 64}, 105007 (2001).
%
%
\bibitem{Gavoret64} J. Gavoret and P. Nozi\`{e}res, {Ann. Phys.} {\bf 28}, 349 (1964).
%
\bibitem{Ledowski04} S. Ledowski, N. Hasselmann, and P. Kopietz, {Phys. Rev.} A {\bf 69}, 061601(R) (2004); N. Hasselmann, S. Ledowski, and P. Kopietz, {\em ibid.} {\bf 70}, 063621 (2004).
%
\bibitem{Blaizot06} J.-P. Blaizot, R. M\'{e}ndez-Galain, and N. Wschebor, 
Phys.~Rev.~E {\bf 74}, 051116 (2006); {\it ibid.}  {\bf 74}, 051117 (2006).
%
\bibitem{Vidberg77} H. J. Vidberg and J. W. Serene, {J. Low Temp. Phys.} {\bf 29}, 179 (1977).
%
\bibitem{Kreisel08} A. Kreisel, F. Sauli, N. Hasselmann, and P. Kopietz, 
{Phys. Rev.} B {\bf 78}, 035127
(2008).
\end{thebibliography}
\end{document}